\documentclass[runningheads]{llncs}

\usepackage[T1]{fontenc}
\usepackage{pdfpages}
\usepackage{hyperref}
\usepackage{url}
\usepackage{graphicx}  
\usepackage{adjustbox} 
\usepackage{caption}   
\usepackage{soul}
\usepackage[inline]{enumitem} 
\usepackage{bm,upgreek}
\usepackage[english]{babel}
\usepackage{booktabs}
\usepackage[controls]{animate}
\usepackage{tikz}
\usetikzlibrary{arrows}
\usepackage{todonotes}
\usepackage{algorithm}
\usepackage{algpseudocode}
\usepackage{amsmath}
\usepackage{amssymb}
\usepackage{amsfonts}
\usepackage{cleveref}
\usepackage{multirow}
\usepackage{subcaption}
\usepackage[utf8]{inputenc} 
\usepackage[T1]{fontenc}    
\usepackage{url}            
\usepackage{booktabs}       
\usepackage{amsfonts}       
\usepackage{nicefrac}       
\usepackage{microtype}      
\usepackage{xcolor}         
\usepackage{orcidlink}
\newcommand{\bfA}{{\bf A}}

\newcommand{\bfx}{{\bf x}}

\newcommand{\bfd}{{\bf d}}
\newcommand{\bfm}{{\bf m}}

\newcommand{\bfz}{{\bf z}}

\newcommand{\bfPsi}{{\bm{\mathrm{\Psi}}}}

\begin{document}

\title{Inversion of Magnetic Data using Learned Dictionaries and Scale Space}

\author{Shadab Ahamed\inst{1}\thanks{Equal contribution}$^\dagger$\orcidlink{0000-0002-2051-6085}\and 
    Simon Ghyselincks\inst{1}$^\star$\orcidlink{0009-0004-9555-0634} \and 
    Pablo Chang Huang Arias\inst{2} \and
    Julian Kloiber\inst{1} \and 
    Yasin Ranjbar\inst{3} \and 
    Jingrong Tang\inst{3} \and 
    Niloufar Zakariaei\inst{2} 
    \and Eldad Haber\inst{2}}

\authorrunning{S. Ahamed et al.}

\institute{Department of Physics \& Astronomy, University of British Columbia, Vancouver, BC, Canada \and
Department of Earth, Ocean and Atmospheric Sciences, University of British Columbia, Vancouver, BC, Canada \and 
Department of Mechanical Engineering, University of British Columbia, Vancouver, BC, Canada}

\maketitle             

\footnotetext[4]{\email{shadab.ahamed@hotmail.com}}

\begin{abstract}
Magnetic data inversion is an important tool in geophysics, used to infer subsurface magnetic susceptibility distributions from surface magnetic field measurements. This inverse problem is inherently ill-posed, characterized by non-unique solutions, depth ambiguity, and sensitivity to noise. Traditional inversion approaches rely on predefined regularization techniques to stabilize solutions, limiting their adaptability to complex or diverse geological scenarios. In this study, we propose an approach that integrates variable dictionary learning and scale-space methods to address these challenges. Our method employs learned dictionaries, allowing for adaptive representation of complex subsurface features that are difficult to capture with predefined bases. Additionally, we extend classical variational inversion by incorporating multi-scale representations through a scale-space framework, enabling the progressive introduction of structural detail while mitigating overfitting. We implement both fixed and dynamic dictionary learning techniques, with the latter introducing iteration-dependent dictionaries for enhanced flexibility. Using a synthetic dataset to simulate geological scenarios, we demonstrate significant improvements in reconstruction accuracy and robustness compared to conventional variational and dictionary-based methods. Our results highlight the potential of learned dictionaries, especially when coupled with scale-space dynamics, to improve model recovery and noise handling. These findings underscore the promise of our data-driven approach for advance magnetic data inversion and its applications in geophysical exploration, environmental assessment, and mineral prospecting. The code is publicly available at: \url{https://github.com/ahxmeds/magnetic-inversion-dictionary.git}.

\end{abstract}

\section{Introduction}
\label{sec:introduction}

Magnetic data is commonly collected to understand the Earth's subsurface properties. Such data is abundant \cite{adagunodo2015overview} and covers extensive areas of the Earth's surface. A key method for interpreting the data is magnetic inversion, 
which aims to reconstruct subsurface magnetic properties, such as susceptibility, from observed magnetic fields \cite{li19963}. It has been a subject of extensive research in geophysics with applications in subsurface characterization, mineral exploration, and environmental research \cite{ellis2012inversion,lelievre2003forward,shi2023deep}.

However, magnetic inversion is a classic example of a highly ill-posed inverse problem, where multiple subsurface models can equally explain the same observed magnetic anomalies. Magnetic signals attenuate with increasing depth, thereby limiting the resolution of deeper structures and compounding the ambiguity \cite{utsugi20193}. Noise in the data and variability in magnetization properties, such as induced and remanent components, can significantly distort the inversion outcomes \cite{ellis2012inversion}. To address these challenges, classical inversion methods incorporate regularization, such as $\ell_1$-$\ell_2$ norms, to balance sparsity and smoothness \cite{li19963,utsugi20193}. Probabilistic frameworks, including linear stochastic inversion introduced by Franklin \cite{franklin1970well} and refined by Tarantola and Valette \cite{tarantola1982generalized,shamsipour20113d}, initially provided a foundation for addressing inversion ambiguities. Building on this probabilistic approach, Bosch et al. \cite{bosch2001joint,bosch2006joint} used Monte Carlo methods for gravity inversion to derive posterior probability densities, capturing the range of plausible models consistent with the data. Chasseriau and Chouteau \cite{chasseriau20033d} emphasized integrating prior geological information through three-dimensional gravity data inversion constrained by an a priori covariance model. Incorporating spatial correlations of subsurface properties enhances stability and geological realism, overcoming the limitations of traditional deterministic methods. Nonetheless, these conventional approaches rely on pre-determined regularization schemes, which may limit their adaptability to complex geological scenarios. More recently, work in the related field of imaging inverse problems has demonstrated the effectiveness of multiscale convolutional dictionary learning as a method for solving challenging problems such as medical image reconstruction \cite{liu2022learning}.

In this work, we propose a similar approach for magnetic inversion by integrating dictionary learning \cite{horesh2007overcomplete} with a scale-space framework. Our framework aims to learn a dynamic dictionary that evolves over iterations, creating a versatile inversion process that can approximate complex models which fixed dictionaries often fail to represent. Numerical experiments demonstrate this method not only outperforms classical methods in terms of accuracy and robustness but also offers greater flexibility in handling diverse geological structures. This analysis highlights the promise of a scale-space approach as an alternative for standard dictionary learning in geophysical inversion.

\section{Overview of the Problem}
\label{sec:overview_of_the_problem}

Magnetic inversion aims to reconstruct a three-dimensional magnetic susceptibility model from two-dimensional surface magnetic data. According to Gauss’s theorem, infinitely many non-unique source distributions can produce the same observed magnetic field. To address these challenges, inversion methods incorporate prior information and regularization strategies to ensure physically plausible and geologically meaningful solutions.

This section introduces the mathematical framework of magnetic inversion, beginning with the forward problem, which establishes the relationship between magnetic susceptibility and observed data. We then discuss the inverse problem and the strategies used to recover a subsurface model. Finally, we explore the variational and scale-space approaches, highlighting their strengths and limitations in addressing the challenges of magnetic inversion. In particular, we focus on dictionary-based inversion techniques, which we subsequently reformulate into a trainable dictionary framework.

\subsection{Forward Problem}
\label{subsec:forward_problem}

The forward problem describes how a given magnetic susceptibility distribution gives rise to an observed magnetic field. It establishes a mathematical relationship between the subsurface model and the measured data, typically expressed as the linear system (Fredholm integral equation of the first kind \cite{tricomi1985integral}) of the form
\begin{equation} \label{eq:Greens}
    d(r) = \int_{\Omega} G(r, r') m(r') \, dr',
\end{equation}
where \( d(r) \) is the magnetic field response, \( r \) and \( r' \) are the observation and source coordinates, respectively, \( \Omega \) is the volume of the 3D susceptibility model, \( G(r, r') \) is the Green's tensor capturing the physical interaction between the source and observation points, and \( m \) is the magnetic susceptibility distribution. For the explicit form of $G$, see \cite{abubakar2004iterative}.
Discretizing the integral, we obtain a linear system of equations,
\begin{equation}
\label{eq:forward-problem}
    \bfd = \bfA \bfm, 
\end{equation}
where \( \bfd \) is the observed magnetic field data, \( \bfA \) is the forward operator (or kernel), and \( \bfm \) is the discrete susceptibility values within subsurface cells \cite{li19963}.

\subsection{Inverse Problem} 
\label{subsec:inverse_problem}

The inverse problem seeks to recover the true subsurface model $\bfm$ from observed magnetic data $\bfd$, given the forward operator $\bfA$, defined in Eq.~\ref{eq:forward-problem}. Due to the ill-posed nature of magnetic inversion, additional constraints and regularization techniques incorporating prior knowledge are necessary to obtain physically plausible and stable solutions. Our work focuses on leveraging scale-space methods to tackle the ambiguities inherent in this inverse problem. We demonstrate the effectiveness of these methods by comparing them with established variational approaches, highlighting their potential to improve accuracy.

\subsubsection{Variational Approach.} 
\label{subsubsec:variational-approach}
The classical variational inversion approach employs a Bayesian statistical framework to incorporate prior information and regularization into the inverse problem \cite{Stuart2010}. We assume that the noise in each observation of $\bfd$ is independent and identically distributed as a zero-mean Gaussian random variable with a diagonal covariance matrix \cite{li19963}. We now discuss two different regularization types that are commonly used. One that is based on analysis and the second that is based on synthesis \cite{selesnick2009signal}
\begin{enumerate}[label=(\roman*)]
\item {\bf Analysis-based regularization:}
To encode prior expectations about the model $\bfm$, we introduce a regularization function $R(\bfm)$. Such a prior is also obtained when considering the Maximum A Posteriori (MAP) estimate. It seeks the  model $\bfm$ that maximizes the posterior, given the observations $\bfd$ and the prior, leading to the following optimization problem or loss function \cite[Section 3.5.6]{zhdanov2002,Stuart2010}:
\begin{equation}
\label{eq:variational-loss-fn}
\phi(\bfm) =  \frac{1}{2} \|\bfA \bfm - \bfd\|^2 + \alpha R(\bfm),
\end{equation}
where $\phi(\bfm)$ is the objective function to be minimized.
The regularization $R(\bfm)$ serves to stabilize the inversion by incorporating prior knowledge or enforcing certain properties in the solution.
The regularization operator is often referred to as the analysis term, as it analyzes potential $\bfm$ vectors. If $R(\bfm)$ is large, then $\bfm$ is deemed to be unlikely. 
In this work, we used a depth-weighted regularization proposed in \cite{li19963} and a total variation approach proposed in \cite{Vogel} as a baseline. In our experience, they achieve very similar results to other popular regularization techniques such as the one proposed in \cite{zhdanov2002}.

\item {\bf Synthesis-Based Regularization:}
A different approach for the regularization of the problem is synthesis or sparse recovery.
We assume the existence of a so-called ``dictionary'' $\bfPsi$ using which the model $\bfm$ is then ``synthesized'' from an unknown vector $\bfz$ by setting,
\begin{equation}
\label{synth}
\bfm = \bfPsi \bfz.
\end{equation}
If the dictionary is chosen appropriately, drawing from signal processing, we can impose a prior belief that $\bfm$ can be represented as a sparse combination of fundamental or atomic distributions \cite{marques,tosic,tillmann} which leads to the following optimization problem,    
\begin{equation}
\label{eq:scale-space}
    \min_\bfz  \frac{1}{2} || \bfA \bfPsi \bfz - \mathbf{d} ||^2 + \alpha || \bfz ||_1. 
\end{equation}
\end{enumerate}

\subsubsection{Inverse Scale-Space Approach.}
\label{subsubsec:scale-space-approach}

The inverse scale-space approach extends the classical variational method by incorporating multi-scale representations, enabling the analysis of structures at varying resolutions \cite{Burger,Worrall}. Rather than solving the optimization problems in Eqs. \eqref{eq:scale-space} or \eqref{eq:variational-loss-fn}, an iterative process is generated such that 
the data is gradually fitted and structure is introduced into the solution. The process is terminated when the model sufficiently fits the data.

Such a process is described in detail for the analysis approach in \cite{Burger}. Here, we extend this to the synthesis approach in a straightforward way. To this end, consider an iterative approach for the solution of the optimization problem Eq. \ref{eq:scale-space} using a soft-thresholding algorithm: 
\begin{equation} 
\label{eq:soft-threshold}
    \bfz^{(j+1)} = \sigma_{\tau} \left(\bfz^{(j)} - \mu_j \bfPsi^\top \bfA^\top \left(\bfA \bfPsi  \bfz^{(j)} -\mathbf{d}\right)\right), 
\end{equation}
where $\mu_j$ is the step-size, $\bfz^{(j)}$ represents the value of $\bfz$ at $j^\text{th}$ iteration and $\tau$ is the thresholding parameter. This approach is one of the most popular for solving the optimization problems in Eq. \eqref{eq:scale-space} and is thoroughly discussed in \cite{donoho1995noising}.\\

In this process, each iteration refines the solution by updating it based on the gradient of the data misfit, followed by the application of the soft-thresholding operator
\begin{equation}
    \sigma_\tau(z_i) = \text{sign}(z_i) \cdot \max \{|z_i| - \tau ,0\}, 
\end{equation}
where, $z_i$ represents the $i^\text{th}$ component of $\bfz$.
The soft-thresholding operator reduces the magnitude of coefficients by a threshold \(\tau\), setting small values to zero, thereby enforcing sparsity and enhancing the stability and robustness of the solution \cite{marques,daubechies}. Here, unlike traditional approaches that iterate to full convergence, we use scale-space methods that terminate early, capturing dominant structures while avoiding overfitting \cite{Burger}. 
Next, we show how to use the scale-space formulation of sparse recovery in the context of dictionary learning for magnetic inversion.

\section{Learning Scale Space}
\label{sec:method}

In the previous section, we explored regularization techniques that are based on variational as well as scale-space approaches. A unifying feature of these approaches is that the regularization is determined a-priori using some ideas about the desired properties of the solution.\\

In this study, we propose a new data-driven approach. We assume to have training data that is a set of plausible models, ${\cal M} = \{\bfm_1,\ldots,\bfm_N \}$. These represent different Earth models that are sampled from some distribution of Earth geologies, $\pi_{\bfm}$. We assume that the models are sampled independently, that is, the set ${\cal M}$ is independent and identically-distributed.
We aim to reconstruct an ``optimal'' regularization in some sense (discussed next). 
In this work, we focus on learning dictionaries for the synthesis approach. Previous studies utilized a predefined \(\bfPsi\), typically by using a cosine or a wavelet transform \cite{yang2011multitask}. 

We propose to use a linear operator, \(\bfPsi\), that is a set of convolutional operations that can be learned, given some training data. For the problem at hand, \(\bfPsi\) is a 4D tensor of dimensions $C \times H \times W \times D$, where $H$, $W$, $D$ denote the height, width and depth of the kernel and $C$ denotes the channels. Formally, we can write $\bfPsi$ as
\begin{equation}
    \label{psi}
    \bfPsi = \Big(\bfPsi_1,  \ldots , \bfPsi_C\Big),
\end{equation}
where $\bfPsi_j$ is a circular convolution matrix, that can be expressed as a stencil of $H \times W \times D$ entries,  $C$ is the number of stencils (or channels). The vector $\bfz = \Big(\bfz_1^{\top} \ldots  \bfz_c^{\top}\Big)^\top$ represents the unknown state that we solve for.

Our goal is to learn $\bfPsi$ given the data set ${\cal M}$. To this end, we use the forward problem and generate the synthetic data $\bfd_j = \bfA \bfm_j + \bm{\epsilon}_j$. Consider the recovery of $\bfm_j$ from $\bfd_j$ given a known (assumed) $\bfPsi$. Assume that we use Eq. \eqref{eq:soft-threshold} to solve for $\bfz$ given $\bfPsi$ and let $\bfz^{(N)}$ be the state that is obtained after $N$ iterations. Furthermore, we let
\begin{equation}
    \label{mhat}
    \widehat \bfm = \bfPsi \bfz^{(N)}, \quad\quad \widehat \bfd = \bfA \widehat \bfm,
\end{equation}
and define the loss
\begin{equation}
\label{eq:loss-fcn}
    {\ell}(\bfPsi) = \frac 12{\mathbb E}_{\bfm}\Big[\|\widehat \bfm - \bfm \|^2\Big] +  \alpha \frac 12{\mathbb E}_{\bfd}\Big[\|\widehat \bfd - \bfd \|^2\Big],
\end{equation}
where $\alpha$ is a hyper-parameter that weighs the contribution of the forward-data misfit term relative to the dictionary-based model misfit. The additional loss term on the forward data promotes solutions that match the observed data from the subset of possible $\widehat \bfm$ which may have the same $\ell_2$-norm distance from the true model. In our study, we used $\alpha=1$ for convenience. Hence, the optimal $\bfPsi$ is the one that minimizes the mean loss over all models in our data set as well as the one that fits to the observed data by minimizing over the mean loss over all data.

At this point, it is interesting to recognize a direct connection between deep learning of convolution neural networks and the proposed method, motivating an examination of Eq.~\eqref{eq:soft-threshold} from the point of view of a neural network. The dictionary $\bfPsi$ is a convolution operation and so is $\bfPsi^{\top}$. The network uses the convolutions together with the forward problem $\bfA$ to form a layer. Finally, the soft-thresholding operator $\sigma_{\tau}$ plays the role of non-linearity. Thus, one can use both theory as well as common software tools in practical implementation of deep learning to solve our problem.

One standard approach for solving such a problem relies on stochastic gradient descent \cite{bottou2012stochastic}. In this approach, we sample the model space ${\cal M}$, use the forward modeling to simulate the data and then use the network
\eqref{eq:soft-threshold} to recover a solution and evaluate the loss. We then use automatic differentiation to compute the gradient of the loss with respect to $\bfPsi$ and take a step in the direction of negative gradient. The method and algorithm is summarized in subsections \ref{subsec:Fixed-Dictionary} and \ref{subsec:Variable-Dictionary} below where we have experimented with two types of dictionary for the scale-space methods: a single shared dictionary and a dynamic iteration-dependent dictionary that is sometimes also referred to as the unrolled dictionary.


\subsection{Shared $\bfPsi$ Dictionary}
\label{subsec:Fixed-Dictionary}

The optimization problem in Eq. \eqref{eq:scale-space} and the corresponding iterative method in Eq. \eqref{eq:soft-threshold} to solve it, uses a single dictionary $\bfPsi$, shown in Algorithm \ref{alg:shared-dict}. In the context of neural networks, the dictionary is a convolution that is shared between the different layers. The advantages of such a convolution is simplicity and robustness. Indeed, the method can be easily interpreted as dictionary learning for the $\ell_1$ recovery problem previously discussed in \cite{horesh2007overcomplete}.
 However, while this approach simplifies computations, it limits the model’s ability to adapt to complex and dynamic patterns in the data.

\begin{algorithm}
\caption{Single Shared Dictionary Learning} \label{alg:shared-dict}
\begin{algorithmic}
\Require Data $\mathcal{M} = \{\bfm_1, \ldots, \bfm_N \}$, operator $\bfA$, learning rate $\alpha$, iterations $N$
\State Initialize a single dictionary $\bfPsi$ of size $1\times C \times H \times W \times D$
\While{not converged}
    \State Randomly sample $\bfm$ from $\mathcal{M}$
    \State Compute $\bfd = \bfA \bfm + \epsilon$
    \State $\bfz^{(0)} \gets \mathbf{0}$ \Comment{Initialize latent representation}
    \For{$j=0$ to $N-1$}
        \State \(
        \bfz^{(j+1)} = \sigma_{\tau} \left(\bfz^{(j)} - \mu_j \bfPsi^\top \bfA^\top \left(\bfA \bfPsi  \bfz^{(j)} -\bfd\right)\right)
        \) \Comment{Soft-thresholding Eq.~\eqref{eq:soft-threshold}}
    \EndFor
    \State Compute $\widehat{\bfm} = \bfPsi\, \bfz^{(N)}$ and $\widehat{\bfd} = \bfA\,\widehat{\bfm}$ \Comment{Reconstruction step Eq.~\eqref{mhat}}
    \State Estimate the loss $\ell =\frac 12 \|\widehat{\bfm} - \bfm\|^2 + \frac 12 \|\widehat{\bfd} - \bfd\|^2$ \Comment{Loss function Eq.~\eqref{eq:loss-fcn}}
    \State Compute $\delta \bfPsi = \nabla_{\bfPsi}\,\ell$ \Comment{Automatic differentiation}
    \State Update $\bfPsi \gets \bfPsi - \alpha \delta \bfPsi$
\EndWhile
\end{algorithmic}
\end{algorithm}

\subsection{Unrolled $\bfPsi$ Dictionary}
\label{subsec:Variable-Dictionary}
A less restrictive approach uses an iteration-dependent dictionary, shown in Algorithm \ref{alg:scalespace-dict}. This allows the dictionary to adapt dynamically throughout the training process, yielding further degrees of freedom. In this case, each layer is assigned its own dictionary \(\bfPsi_{(j)}\), allowing the model to better capture intricate and evolving patterns in the data. This is achieved by using a separate set of weights for each iteration, unlike the fixed-dictionary approach where weights are shared.
However, because each iteration has its own dictionary, we lose a single global objective in the strict variational sense - there is no single functional whose minimizer the algorithm is guaranteed to converge to. Nonetheless, this is exactly the power of scale-space methods which introduce a dynamical process for the regularization without it having to be rooted in optimization.
The iterative method for the case of unrolled dictionary can be written as,
\begin{equation}
\label{eq:vardict}
\bfz^{(j+1)} = \sigma_{\tau} \left(\bfz^{(j)} - \mu_j \bfPsi_{(j)}^\top \bfA^\top \left(\bfA \bfPsi_{(j)} \bfz^{(j)} -\mathbf{d}\right)\right).
\end{equation}
Note that it requires updating $N$ different dictionaries rather than a single one. It has been shown in \cite{monga2021algorithm} that such relaxation regularization can yield better recoveries compared to fixed ones.

\begin{algorithm}
\caption{\textbf{Unrolled Dictionary Learning}} \label{alg:scalespace-dict}
\begin{algorithmic}
\Require Data $\mathcal{M} = \{\bfm_1, \ldots, \bfm_N \}$, operator $\bfA$, learning rate $\alpha$, iterations $N$
\State Initialize a set of dictionaries $\bfPsi_{(1)},...,\bfPsi_{(N)}$ of size $N\times C \times H \times W \times D$
\While{not converged}
    \State Randomly sample $\bfm$ from $\mathcal{M}$
    \State Compute $\bfd = \bfA \bfm + \epsilon$
    \State $\bfz^{(0)} \gets \mathbf{0}$ \Comment{Initialize latent representation}
    \For{$j=0$ to $N-1$}
        \State \(
        \bfz^{(j+1)} = \sigma_\tau \left(\bfz^{(j)} - \mu_j \bfPsi_{(j)}^\top \bfA^\top \left(\bfA \bfPsi_{(j)} \bfz^{(j)} -\mathbf{d}\right)\right)
        \) \Comment{Soft-thresholding Eq.~\eqref{eq:vardict}}
    \EndFor
    \State Compute $\widehat{\bfm} = \bfPsi\, \bfz^{(N)}$ and $\widehat{\bfd} = \bfA\,\widehat{\bfm}$ \Comment{Reconstruction step Eq.~\eqref{mhat}}
    \State Estimate the loss $\ell =\frac 12 \|\widehat{\bfm} - \bfm\|^2 + \frac 12 \|\widehat{\bfd} - \bfd\|^2$ \Comment{Loss function Eq.~\eqref{eq:loss-fcn}}
    \State Compute $\delta \bfPsi_{(j)} = \nabla_{\bfPsi_{(j)}}\,\ell$ for all $j$
    \State Update $\bfPsi_{(j)} \gets \bfPsi_{(j)} - \alpha \delta \bfPsi_{(j)}$ for all $j$
\EndWhile
\end{algorithmic}
\end{algorithm}

\section{Numerical Experiments}
\label{sec:numerical_experiments}

In this section, we evaluate experiments in 3D that demonstrate the advantage of learning regularization over pre-determined models. We used synthetic data and experimented with four different methods, (i) a variational approach that is based on the gradients of the model, (ii) a sparse recovery with pre-determined fixed dictionary that is based on the cosine transform, (iii) a shared learned dictionary that approximately minimizes Eq. \eqref{eq:scale-space}, and (iv) a learned unrolled dictionary that is iteration-dependent (Eq. \eqref{eq:vardict}). The first two approaches were used as baselines to the learned approaches proposed in this work. The code to reproduce our results is publicly available at: \url{https://github.com/ahxmeds/magnetic-inversion-dictionary.git}.

\subsection{Dataset}
We performed experiments using a $64\times64\times32$ uniform grid of $100\text{m}^3$ cells with a baseline magnetic susceptibility of $\chi_v = 0$. A randomly selected number of features were added to the model to form ellipsoid shaped areas of increased magnetic susceptibility with exponential radial decay (see Table~\ref{tab:dataset}). Such models simulate the existence of mineral deposits that are commonly present in geological structures such as porphyries \cite{sinclair2007porphyry}. Given these models, we train the dictionary(ies) $\bfPsi$ with a training set of 2000 samples and a validation set of 100 samples. The model with the lowest loss on the validation set was used for final testing on the unseen test set consisting of 500 samples. We added $1\%$ relative noise to the forward data to simulate measurement errors before performing the inversion. 
\begin{table}[tb]
\centering
\caption{Summary of the dataset parameters. \(\mathcal{U}\) indicates a uniform distribution.}
\label{tab:dataset}
\begin{tabular}{l @{\hspace{1em}} l @{\hspace{2em}} l @{\hspace{1em}}l}
\toprule
\textbf{Parameter} & \textbf{Value} & \textbf{Parameter} & \textbf{Value} \\ 
\midrule
Grid dimensions & \(64 \times 64 \times 32\) & Feature count & \(n\sim \{1,\dots,6\}\) \\
Feature origin & $\bfx_0 \sim \mathcal{U}[0.2, 0.8]^3$ & Amplitude & $\chi_0 \sim \mathcal{U}[0,1]$ \\
Susceptibility decay & $\chi_v = \chi_0 \exp(-50 \| \mathbf{x}-\mathbf{x}_0 \|^2)$ & Dataset size & 2600 samples \\
\bottomrule
\end{tabular}
\end{table}

\subsection{Method Specifications}
We now discuss the details for each of the methods use for the recovery.

\subsubsection{Variational:}
While variational approaches are pre-determined, they involve a number of hyperparameters. The optimal regularization hyperparameters for the variational approach in Eq.~\ref{eq:variational-loss-fn} were determined via grid search method, minimizing the $\ell_2$ loss between the predicted and true model, $\|\hat \bfm - \bfm\|^2$. 

\subsubsection{Cosine $\bfPsi$:}
For the cosine based dictionary, we choose the truncation and the step length based on the work proposed in \cite{rubinstein2010dictionaries}. 

\subsubsection{Learned $\bfPsi$:}
Both the shared $\bfPsi$ and the unrolled $\bfPsi^{(i)}$ used a 3D convolution of kernel dimensions $7 \times 7 \times 7$ with a channel dimension $C=9$ and a fixed total of $N=16$ iterations in Algorithms \ref{alg:shared-dict} and \ref{alg:scalespace-dict} for direct comparison. The unrolled dictionary had a new set of weights on each iteration adding complexity compared to the single learned dictionary of shared weights.

\subsection{Comparison of Model Recovery}

\begin{figure}[tb]
    \centering
    \includegraphics[width=0.48\linewidth]{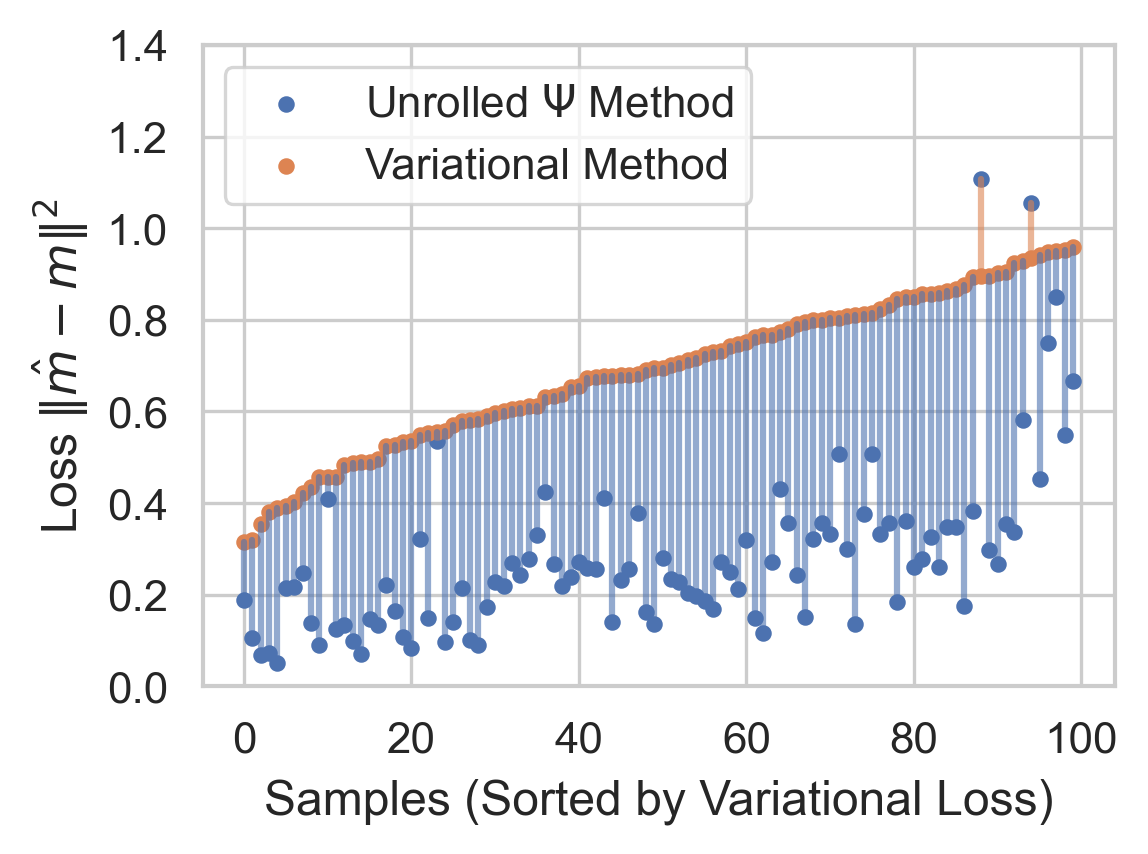}
    \hfill
    \includegraphics[width=0.48\linewidth]{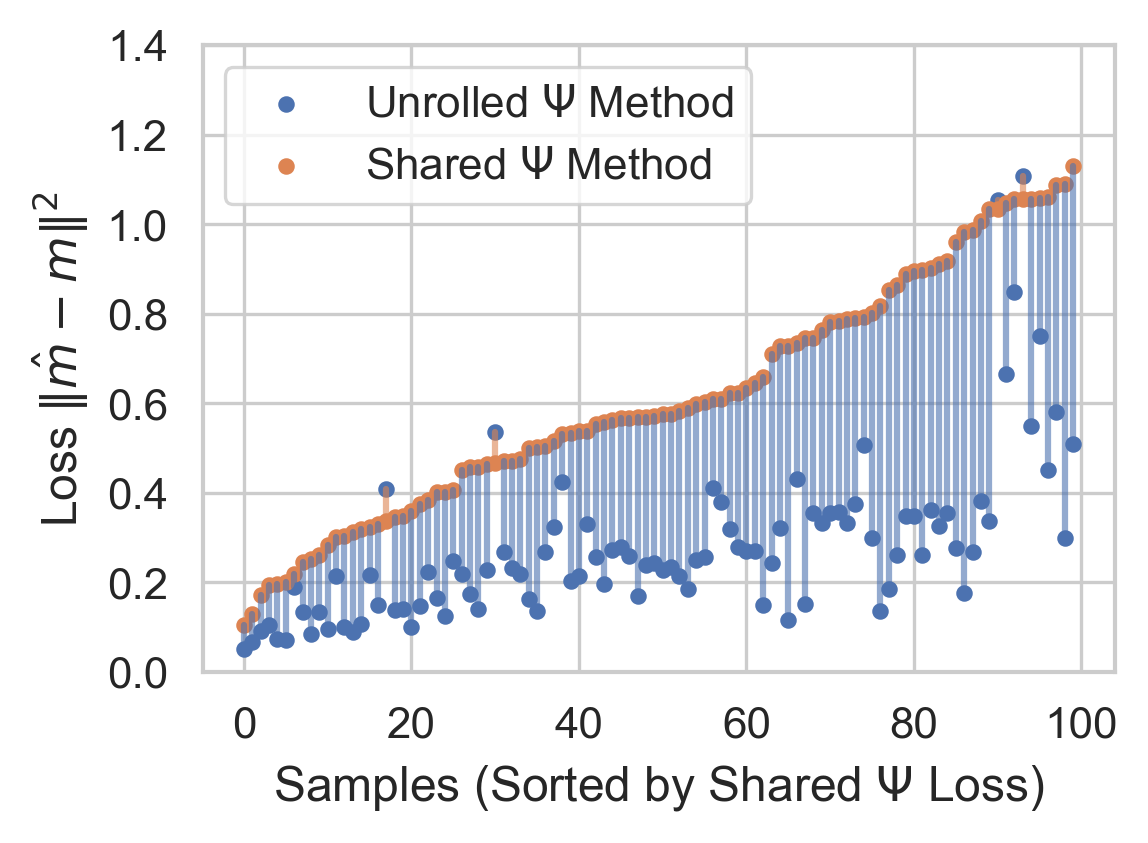}
    \caption{Comparison of loss values obtained from 100 samples randomly drawn from the test set using different methods. The left panel compares the variational method with the unrolled $\bfPsi$ method, while the right panel compares the single shared $\bfPsi$ method with the unrolled $\bfPsi$ method. In both cases, the unrolled $\bfPsi$ (blue) consistently outperformed the competing models (orange) based on the nMSE loss for model recovery.}
    \label{fig:side-by-side-loss-comparison}
\end{figure}

\begin{figure}[ht]
    \centering
    \includegraphics[width=0.9\linewidth]{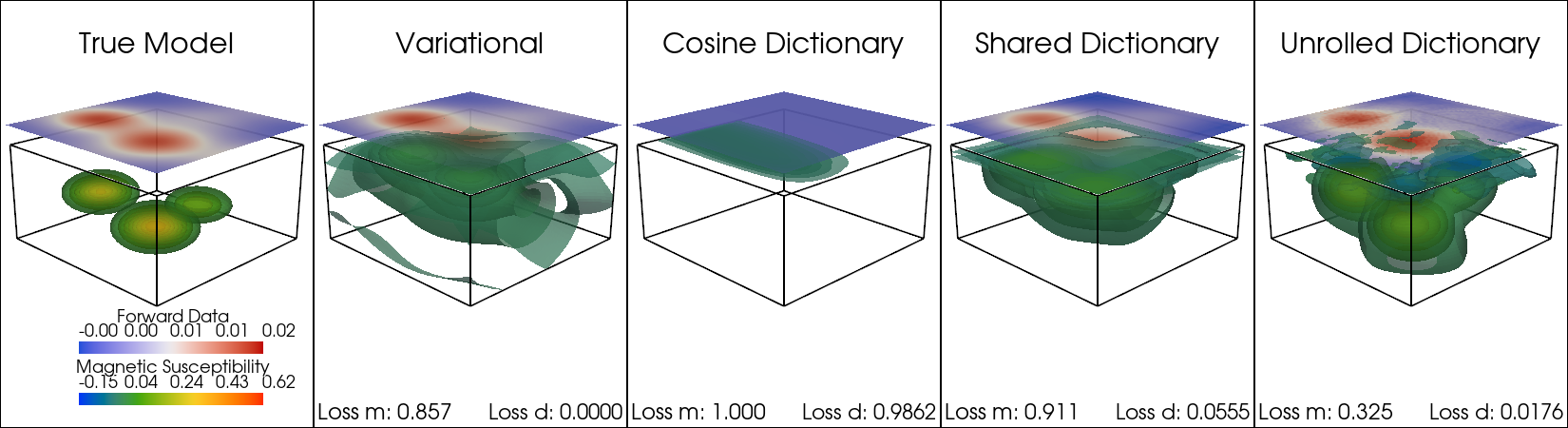}
    \vspace{0.3em} 

    \includegraphics[width=0.9\linewidth]{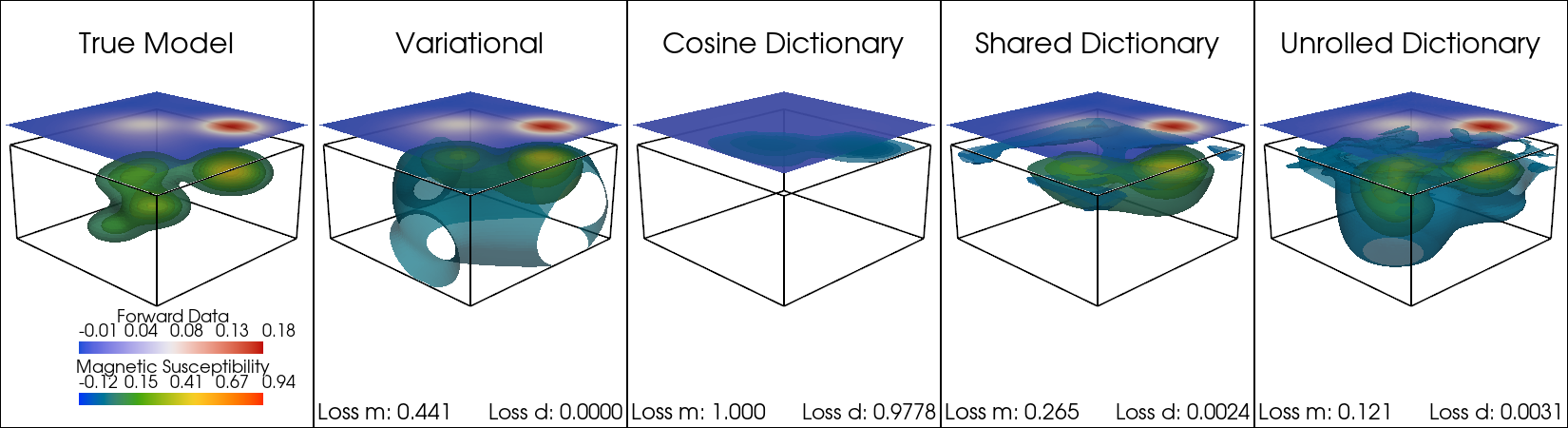}
    \vspace{0.3em} 

    \includegraphics[width=0.9\linewidth]{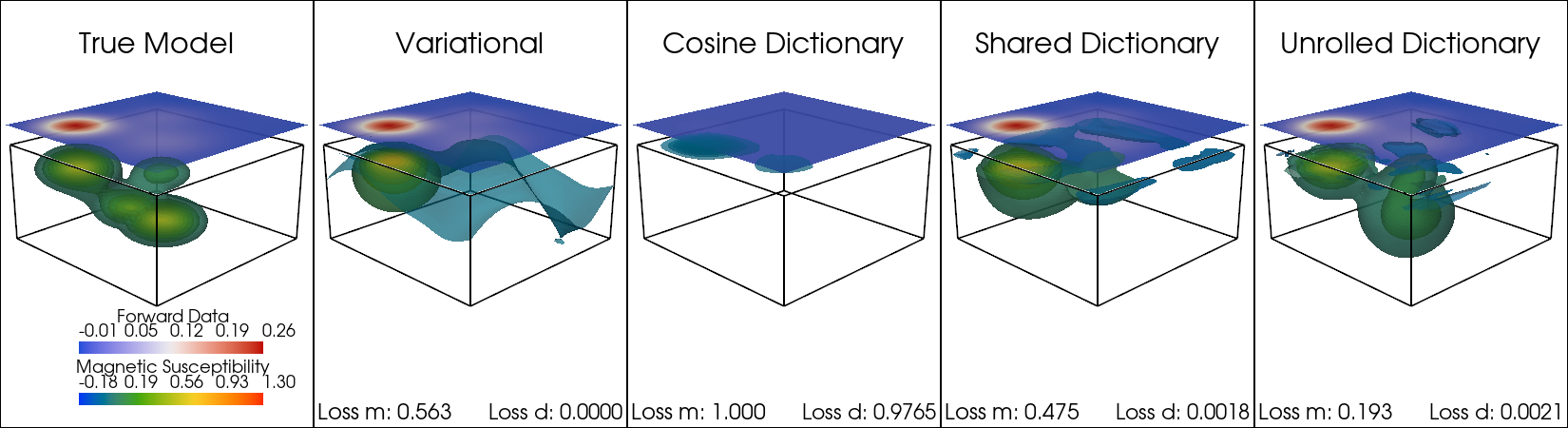}

    \caption{Three sample models (01, 11, and 19) from the test set illustrating the quality improvement of unrolled dictionary learning for inversion compared to other methods. Each image highlights different features of the model reconstruction quality achieved using the proposed method. The 2D magnetic data from the 3D true model is inverted back to a 3D construction using four different methods. Views of the reconstruction are shown using contour potentials of the magnetic susceptibility.}
    \label{fig:stacked-3d-model-comparison}
\end{figure}

A comparison of the normalized mean squared error (nMSE) between the four different methods is shown in Table~\ref{tab:nmse_results}. Qualitative comparison between different methods for the recovery of 3D models has been shown for some representative test set samples in Fig.~\ref{fig:stacked-3d-model-comparison}. In addition, we compare the distribution of losses between variational methods and unrolled $\bfPsi$ in Fig.~\ref{fig:side-by-side-loss-comparison} (left) and between shared $\bfPsi$ and unrolled $\bfPsi$ in Fig.~\ref{fig:side-by-side-loss-comparison} (right).

\begin{table}[]
\centering
\caption{Normalized Mean Squared Error (nMSE) for model and data recovery using different methods. \textbf{Bold} and \textit{italicized} values indicate the best and the second best method for each recovery type.}
\label{tab:nmse_results}
\resizebox{0.9\columnwidth}{!}{%
\begin{tabular}{@{}llclc@{}}
\toprule
\multicolumn{1}{c}{\textbf{Recovery}} & \multicolumn{1}{c}{\textbf{Variational}} & \textbf{\begin{tabular}[c]{@{}c@{}}Cosine \\ $\bfPsi$ \end{tabular}} & \multicolumn{1}{c}{\textbf{\begin{tabular}[c]{@{}c@{}}Shared \\ $\bfPsi$\end{tabular}}} & \textbf{\begin{tabular}[c]{@{}c@{}}Unrolled \\ $\bfPsi$\end{tabular}} \\ \midrule
Model recovery & 0.692{\scriptsize $\pm$0.160}  & 1.000{\scriptsize $\pm$0.000} & \textit{0.625{\scriptsize $\pm$0.250}}  & \textbf{0.289{\scriptsize $\pm$0.161}} \\
Data recovery & \textbf{0.000\scriptsize $\pm$0.000} & 0.983{\scriptsize $\pm$0.006} & 0.032{\scriptsize $\pm$0.077} & \textit{0.021{\scriptsize $\pm$0.071}} \\ \bottomrule
\end{tabular}%
}
\end{table}

The results clearly demonstrate a remarkable improvement of  learnable dictionaries over fixed regularization or dictionaries. The improvement is not marginal. We observe an over $60\%$ improvement in the recovery on average compared to variational approaches and over $55\%$ compared to fixed dictionary.
Comparing the unrolled dictionary that represents a scale space approach to the learned single dictionary in Figure~\ref{fig:stacked-3d-model-comparison}, we observe that the learned variable dictionary produces inverse models that capture features deeper below the subsurface and with closer fidelity to the true model, emphasizing the advantage of scale space methods over variational methods.

\section{Conclusion}
\label{sec:conclusion}

In this paper, we introduced a learning-based approach to dictionary-based inverse problems with an application to the inversion of magnetic data. In particular, we used the scale-space framework and introduced an unrolled version of sparse recovery, where at each iteration, a different dictionary is used. We choose the dictionary based on a convolution that uses multiple channels. We show that this choice can be viewed as a version of deep convolution neural network with a softmax activation. 

A learnable framework requires a training data set. In this work, we assumed the availability of a geostatistical data set that is based on geological modeling for porephyries and trained the dictionaries using simulated data set. Results show that learnable dictionaries significantly outperformed standard techniques. Furthermore, results that are based on unrolled methods which can be thought of as a version of scale-space techniques, outperformed a single shared dictionary. 

It is important to realize that any learnable framework requires a training set which may be challenging to obtain in some scenarios. Further research is required to understand the limitation and the generalization properties of the proposed regularization to models that are out-of-distribution.

\bibliographystyle{splncs04}
\bibliography{references}

\end{document}